\documentclass{article}

\usepackage[final]{neurips_2025}

\usepackage[utf8]{inputenc}
\usepackage[T1]{fontenc}
\usepackage{hyperref}
\usepackage{url}
\usepackage{booktabs}
\usepackage{amsfonts}
\usepackage{amsmath}
\usepackage{microtype}
\usepackage{xcolor}
\usepackage{graphicx}
\usepackage{tikz}
\usetikzlibrary{positioning,arrows.meta,fit,backgrounds,calc}
\usepackage{enumitem}
\setlist{topsep=2pt, itemsep=2pt, parsep=0pt}

\makeatletter
\renewcommand{\@noticestring}{Machine Learning for Systems workshop at the
  40th Conference on Neural Information Processing Systems (NeurIPS 2026).}
\makeatother

\title{Coco: An Agentic Copilot for the\\
  Hardware--Software Co-Design Lifecycle}

\author{%
  \textbf{Samuel Kushnir}$^{*,2}$ \quad
  \textbf{Kavya Sreedhar}$^{*,1}$ \quad
  \textbf{Yeshwanth Reddy Pogula}$^1$ \quad
  \textbf{Amir Yazdan Bakhsh}$^2$ \\
  \textbf{Narges Shahidi}$^1$ \quad
  \textbf{Ming Liu}$^1$ \quad
  \textbf{Varun Gohil}$^3$ \quad
  \textbf{Ravi Iyer}$^1$ \\
  \textbf{Christina Delimitrou}$^3$ \quad
  \textbf{Parthasarathy Ranganathan}$^1$ \quad
  \textbf{Suvinay Subramanian}$^2$ \\
  \vspace{0.2cm}
  $^*$Equal contribution \\
  $^1$Google \quad
  $^2$Google DeepMind \quad
  $^3$MIT
}

\begin{document}

\maketitle

\begin{abstract}
Co-designing ML models and the accelerators that run them is an unusual
reasoning task: architects must draw confident, high-stakes conclusions about
systems that do not yet exist, and the pace of both model evolution and
hardware cadence means the analysis burden grows every quarter. The evidence
behind each decision---hundreds of gigabytes of fresh simulation sweeps over
novel design points---is by construction absent from any LLM's pretraining
corpus. There is no external literature to retrieve; naive
``chat-with-your-data'' approaches hallucinate and lack reliability. We present \textsc{Coco} (\emph{Copilot for Codesign}), an
agentic platform deployed with TPU architects that accelerates the co-design
lifecycle of setting up experiments, sweeping simulators, and deriving
insights. Coco is built as four layers: (i) a \emph{datastore} that
automatically registers every simulation sweep into a normalized relational
schema, so agents ground every number in a SQL query rather than scraping
heterogeneous files; (ii) a library of \emph{tools} with typed APIs that
agents compose without human orchestration; (iii) \emph{agents} that encode
recurring analysis workflows---most notably iso-execution analysis, which
compares systems at matched execution configurations, including swept-but-dominated
points off the Pareto frontier; and (iv) a \emph{platform} UX whose
navigation state doubles as agent context. We report early deployment
experience toward a $2\times$ reduction in time-to-simulation and
time-to-insight, and argue that co-design is a distinct agentic domain:
its data must be retrieved rather than memorized, its workflows are
recurring but context-dependent, and expert adoption hinges on UX that
balances IDE-style control with interactive exploration.
\end{abstract}

\section{Introduction}
\label{sec:intro}

Hardware--software co-design for ML accelerators is a peculiar kind of
performance engineering. First, it reasons about systems that do \emph{not}
yet exist: the artifact under study is a proposed chip, interconnect, or
packaging option, observable only through simulation and analytical models.
Second, its decisions are high-stakes and long-lived---a datacenter
accelerator generation commits billions of dollars years before deployment.
Third, demand is relentless and increasing: hardware cadence is
accelerating, product complexity is growing, and the workload side
diversifies just as fast---dense versus mixture-of-experts
models~\citep{fedus2022switch}, proprietary versus open-source families,
autoregressive versus diffusion decoding.

Concretely, the day-to-day work is dominated by a recurring question shape:
\emph{a number moved---what changed, why, and does it match our intuition?}
When specifications and model assumptions are refreshed each quarter, the
projected advantage of a candidate system over its predecessor can shift by
tens of percent; an architect must decompose that shift into spec changes,
model changes, and sharding changes before anyone acts on it. Comparative
questions have the same texture: why does a dragonfly-class interconnect
benefit decode-heavy inference more than a torus---is it only interconnect
pressure, or also memory bandwidth? Why is model parallelism rarely chosen
on the Pareto frontier? Today, answering any of these places a senior
architect on the critical path for hours-to-days of manual sweep archaeology:
locating the right simulation runs, decoding configuration encodings,
cross-checking simulator quirks, and assembling plots. The experts scale
linearly; the questions do not.

\begin{figure}[t]
\centering
\resizebox{\textwidth}{!}{%
\begin{tikzpicture}[
  font=\small,
  layer/.style={draw, rounded corners=2pt, minimum width=30mm, minimum
                height=7.5mm, align=center},
  side/.style={font=\scriptsize, align=left, anchor=west},
  lft/.style={font=\scriptsize\itshape, align=right, anchor=east},
  stage/.style={draw, rounded corners=2pt, align=center, inner sep=3pt,
                font=\scriptsize, minimum height=7mm},
  arr/.style={-{Stealth[length=2mm]}, thick},
]
% stack
\node[layer, fill=orange!15] (plat) {Platform};
\node[layer, fill=green!12, below=2.5mm of plat] (agents) {Agents};
\node[layer, fill=blue!12, below=2.5mm of agents] (tools) {Tools};
\node[layer, fill=red!12, below=2.5mm of tools] (data) {Datastore};
\node[side, right=3mm of plat] {orchestrate agents with the right UX\\
  \emph{chat, charts, navigation as context}};
\node[side, right=3mm of agents] {automate workflows with apt context\\
  \emph{e.g., iso-execution analysis}};
\node[side, right=3mm of tools] {bite-size tasks behind typed APIs\\
  \emph{composed by agents, not humans}};
\node[side, right=3mm of data] {sweeps auto-registered in a relational DB\\
  \emph{plus specs, hyperparameters, docs}};
\node[lft, left=3mm of plat] {address user intent \ldots};
\node[lft, left=3mm of agents] {\ldots by encoding workflows \ldots};
\node[lft, left=3mm of tools] {\ldots that leverage tools \ldots};
\node[lft, left=3mm of data] {\ldots over a corpus of data};
% lifecycle loop
\begin{scope}[shift={(10.1,-1.55)}]
\node[stage, fill=blue!10] (setup) at (90:1.35) {setup\\experiment};
\node[stage, fill=red!10] (sweep) at (210:1.35) {sweep \&\\simulate};
\node[stage, fill=green!10] (insight) at (-30:1.35) {analyze \&\\derive insights};
\draw[arr] (setup) to[bend right=25] (sweep);
\draw[arr] (sweep) to[bend right=25] (insight);
\draw[arr] (insight) to[bend right=25] (setup);
\node[font=\scriptsize\itshape, align=center] at (0,0) {co-design\\lifecycle};
\end{scope}
\end{tikzpicture}%
}
\caption{Coco's four-layer stack (left) and the co-design lifecycle it
accelerates (right). Each layer is consumed by the one above it; the
platform's navigation state feeds back into agent context.}
\label{fig:stack}
\end{figure}
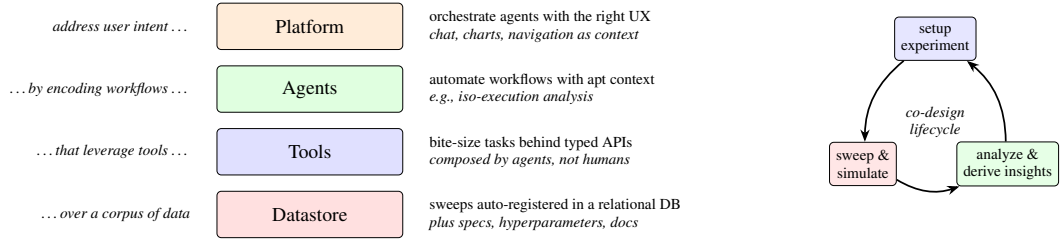

We built \textsc{Coco} (\emph{Copilot for Codesign}), an agentic platform
that accelerates this lifecycle end to end---setting up experiments,
sweeping and simulating, and analyzing the results into
insights (Figure~\ref{fig:stack}, right)---with a near-term goal of halving
time-to-simulation and time-to-insight, and a longer-term goal of letting
the underlying LLM itself iterate on and explore new architectural ideas.
This paper describes the thesis behind Coco (\S\ref{sec:thesis}), its
four-layer architecture (\S\ref{sec:stack}), early deployment experience
with TPU~\citep{jouppi2023tpu} architects (\S\ref{sec:experience}), and two
end-to-end case studies on public TPU generations (\S\ref{sec:cases}).

\section{Why co-design needs its own copilot}
\label{sec:thesis}

Our thesis is that co-design has unique \emph{workflows}, \emph{context},
and \emph{artifacts}, and that a generic coding or chat assistant serves
none of them well.

\textbf{The data cannot be memorized.} Every design cycle regenerates
hundreds of gigabytes of simulation data over design points that are
novel by construction---new chips, new topologies, new model variants.
No pretrained model has seen these numbers, and unlike open-domain
retrieval-augmented generation~\citep{lewis2020retrieval} there is no
external literature to fall back on: the corpus is the organization's own
sweeps, specs, and documents, and it changes under the agent's feet.
Grounding must therefore be \emph{structural}: every quantitative claim
should be traceable to a series of queries and tool calls which a human can easily validate.

\textbf{The workflows recur, but the context varies.} 
Analyses repeat in skeleton but every instance differs in the particulars. Almost every workflow consists of running a sweep, extracting Pareto frontiers, comparing systems, and decomposing a
delta but the particulars: which use case, which sequence-length regime, which metric, are constantly changing. The productive division of
labor is to standardize the skeleton as tools and encoded workflows and let
the LLM handle the \emph{difference}: ``rerun this sweep with only the HBM
bandwidth changed,'' ``pull peak memory usage for the Pareto-optimal points
of that chart.''

\textbf{The artifacts are collaborative objects.} The unit of progress is
not a code diff but a Pareto curve, a per-op breakdown, a utilization
timeline, a comparison report. Experts need to create, navigate, diff, and
share these rapidly---and the trail of artifacts a user has visited is
itself the best statement of their intent.

\section{The Coco stack}
\label{sec:stack}

\paragraph{Datastore: sweeps as a database, not files.} Every simulation
sweep---serving, training, and sampling/thinking workloads, from both
analytical and event-driven simulators---is automatically registered into a
normalized relational schema: test plans, workloads, systems, execution
strategies, Pareto frontiers, and per-phase (prefill/decode/combined)
serving points carrying latency and throughput metrics down to time-to-first-token.
Agents retrieve data by constructing SQL against this schema rather than
scraping CSV, proto, or JSON artifacts. In our experience this is the
single largest lever against hallucination: the model never ``remembers'' a
number, it fetches one, and the query itself is an auditable provenance
record. The datastore also collates the canonical context around the
numbers---system specifications, model hyperparameters, ISA and product
documentation---so retrieval spans both measurements and their meaning.

\paragraph{Tools: bite-size tasks behind typed APIs.} On top of the
datastore sits a library of tools for well-specified tasks, registered with
documentation and made available to LLM agents in the style of modern
tool-use systems~\citep{schick2023toolformer,yao2023react}. A
representative retrieval chain---find valid system/model combinations
$\rightarrow$ find valid Pareto curves $\rightarrow$ fetch a curve
$\rightarrow$ fetch a point $\rightarrow$ fetch its detailed breakdown---is
composed by the agent, interleaved with small amounts of inline code, with
no human orchestration or bespoke glue per analysis.

\paragraph{Agents: encoded workflows with the right context.} Agents
automate the recurring workflows: run an analysis across systems and
models, synthesize the per-system results into a coherent comparison, and
generate a report. The flagship is \emph{iso-execution analysis}, an analysis workflow in which we focus on a given latency target, and directly compare the two systems performance in-depth at that latency target. This includes diving into op-by-op breakdowns and pinning down exactly which operators and under which use cases the systems differ the most on performance. Architects constantly ask why an
under-performing system did not adopt the winning system's configuration
(``why does the torus stop at 16-way expert parallelism when the dragonfly
chooses 64-way?''). The answer usually exists in the sweep---the
configuration was evaluated and dominated by a better execution strategy so it doesn't appear on the Pareto curve. Exhuming it by hand means
reconstructing equivalent configuration keys across systems and locating
which shard of which sweep job produced the point. The agent retrieves the
iso-sharding, iso-batch counterpart directly from the datastore and reports
\emph{why} it lost, turning a multi-hour human search into part of every
report. Coco is now a multi-agent system, where the orchestrator agent can autonomously launch sub-agents with access to the same tooling. In the iso-execution analysis workflow, sub-agents are used to parallelize the analysis across different latency targets and reviewer agents are then used to check over the integrity of the analysis. The orchestrator agent then synthesizes everything to a single report for the human engineer.

\paragraph{Platform: UX as context capture.} The platform provides the
interaction grammar of architectural analysis---hover a Pareto point for
details, click through to its breakdown, open the utilization timeline,
diff two breakdowns across systems or execution strategies---with an
embedded conversational side panel. Critically, the UX is also how agent
context gets captured: the chart the user is looking at, the use case and
sequence-length regime it encodes, and the trajectory of questions that led
here are all attached to the user query automatically, so a user can ask ``why does this point have
poor utilization?'' and the agent will have full understanding of the pareto curve and point in question.

\section{Early experience}
\label{sec:experience}

Coco is deployed with TPU architects and in daily use for serving-focused
analyses; training and further simulator backends are being onboarded.
Three observations from early use:

\textbf{Structured retrieval changes trust.} Analyses that quote numbers
with their generating SQL are treated as grounded evidence by
architects, in a way free-form LLM summaries never were. The schema, not
the model, carries the burden of being right about data.

\textbf{Iso-execution defaults surface real mechanisms.} Comparisons that
previously ended at ``system A wins by $X\%$'' now routinely end at
\emph{which} configurations each system chose and what dominated the
alternatives---the level at which architects can connect results to
intuition, or flag that a simulator assumption (not the hardware) drove the
delta.

\textbf{Context capture is half the product.} The single biggest usability
gap between a generic chatbot and Coco is not model quality but that Coco
knows what the user is looking at. Encoding the navigation trail as agent
context eliminated most of the prompt-writing that made earlier assistants
feel slower than doing the analysis by hand.

Our headline target is a $2\times$ reduction in time-to-simulation and
time-to-insight.

\section{Case studies}
\label{sec:cases}

Two recent analyses exercise the stack end to end. Both involve public TPU
generations: the deployed Ironwood systems and the announced
eighth-generation inference chip, TPU~8i~\citep{google2026tpu8}.

\subsection{Generation-over-generation analysis: Ironwood to TPU 8i}

We tasked Coco with a comparative report between a baseline
Ironwood~\citep{google2025ironwood} system and a (then) next-generation TPU~8i
configuration, asking specifically how the increased on-chip SRAM and
compute change execution. Coco synthesized the simulation sweeps into a
causal, regime-scoped insight rather than a headline ratio: tripling the
scratchpad (128\,MB to 384\,MB) specifically benefits MoE models with
fewer than 64 experts under a sub-2-second latency constraint. It also
explained \emph{why}: at those model shapes, the larger SRAM admits a batch-size
combination at which expert tensors can be pinned and prefetched to fit
the new capacity exactly, taking HBM refetch off the critical path---and
outside that regime (more experts, or looser latency targets) the benefit
washes out. Every quoted number in the report links to its generating SQL.

\subsection{Network topology what-if: torus versus Boardfly}

Ironwood pods connect chips in a torus; TPU~8i moves to a hierarchical
\emph{Boardfly} topology---fully connected four-chip boards, fully
connected eight-board groups, 36 groups per pod~\citep{google2026tpu8}. We
used Coco for a what-if analysis of the two interconnects under highly
sharded MoE serving. Orchestrating the iso-execution tool chain, the agent
held sharding and batch size fixed across topologies to isolate the
network's contribution, showing how Boardfly reshapes the all-to-all
dispatch and combine latencies that dominate expert-parallel execution. It
then generated side-by-side timeline breakdowns from which the architect
could visually verify the bottleneck shifting from network-bound to
compute-bound execution---the level of evidence at which a topology
discussion stops being a matter of taste.

\section{Conclusion}

Hardware--software co-design requires more than generic language models; architects need tools capable of reasoning over fresh simulation data and executing complex, context-dependent workflows. Coco addresses this gap by shifting the burden of data archaeology and routine analysis from the expert to an agentic platform. Rather than relying on model memorization, Coco grounds its insights in a relational datastore and automates recurring tasks like iso-execution comparisons, all within an interface built for active architectural exploration.

While prior work has demonstrated the value of machine learning in physical design~\citep{mirhoseini2021graph}, domain-specific chat assistants~\citep{liu2023chipnemo}, and early co-design productivity studies~\citep{sreedhar2025leveraging}, Coco elevates this assistance directly to the system-architecture level where the highest-stakes decisions are made. By significantly accelerating time-to-insight for architects today, this four-layer stack provides a practical foundation for the autonomous co-designers of tomorrow.

\bibliographystyle{plainnat}
\bibliography{references}

\begin{thebibliography}{10}
\providecommand{\natexlab}[1]{#1}
\providecommand{\url}[1]{\texttt{#1}}
\expandafter\ifx\csname urlstyle\endcsname\relax
  \providecommand{\doi}[1]{doi: #1}\else
  \providecommand{\doi}{doi: \begingroup \urlstyle{rm}\Url}\fi

\bibitem[Fedus et~al.(2022)Fedus, Zoph, and Shazeer]{fedus2022switch}
William Fedus, Barret Zoph, and Noam Shazeer.
\newblock Switch transformers: Scaling to trillion parameter models with simple and efficient sparsity.
\newblock \emph{Journal of Machine Learning Research}, 23\penalty0 (120):\penalty0 1--39, 2022.

\bibitem[{Google}(2026)]{google2026tpu8}
{Google}.
\newblock Our eighth generation {TPUs}: two chips for the agentic era.
\newblock \url{https://blog.google/innovation-and-ai/infrastructure-and-cloud/google-cloud/eighth-generation-tpu-agentic-era/}, 2026.

\bibitem[{Google Cloud}(2025)]{google2025ironwood}
{Google Cloud}.
\newblock Ironwood: The first {Google} {TPU} for the age of inference.
\newblock \url{https://blog.google/products/google-cloud/ironwood-tpu-age-of-inference/}, 2025.

\bibitem[Jouppi et~al.(2023)Jouppi, Kurian, Li, Ma, Nagarajan, Nai, Patil, Subramanian, Swing, Towles, et~al.]{jouppi2023tpu}
Norman~P Jouppi, George Kurian, Sheng Li, Peter Ma, Rahul Nagarajan, Lifeng Nai, Nishant Patil, Suvinay Subramanian, Andy Swing, Brian Towles, et~al.
\newblock {TPU} v4: An optically reconfigurable supercomputer for machine learning with hardware support for embeddings.
\newblock In \emph{Proceedings of the 50th Annual International Symposium on Computer Architecture}, pages 1--14, 2023.

\bibitem[Lewis et~al.(2020)Lewis, Perez, Piktus, Petroni, Karpukhin, Goyal, K{\"u}ttler, Lewis, Yih, Rockt{\"a}schel, et~al.]{lewis2020retrieval}
Patrick Lewis, Ethan Perez, Aleksandra Piktus, Fabio Petroni, Vladimir Karpukhin, Naman Goyal, Heinrich K{\"u}ttler, Mike Lewis, Wen-tau Yih, Tim Rockt{\"a}schel, et~al.
\newblock Retrieval-augmented generation for knowledge-intensive {NLP} tasks.
\newblock \emph{Advances in Neural Information Processing Systems}, 33:\penalty0 9459--9474, 2020.

\bibitem[Liu et~al.(2023)Liu, Ene, Kirby, Cheng, Pinckney, Liang, Alben, Anand, Banerjee, Bayraktaroglu, et~al.]{liu2023chipnemo}
Mingjie Liu, Teodor-Dumitru Ene, Robert Kirby, Chris Cheng, Nathaniel Pinckney, Rongjian Liang, Jonah Alben, Himyanshu Anand, Sanmitra Banerjee, Ismet Bayraktaroglu, et~al.
\newblock {ChipNeMo}: Domain-adapted {LLMs} for chip design.
\newblock \emph{arXiv preprint arXiv:2311.00176}, 2023.

\bibitem[Mirhoseini et~al.(2021)Mirhoseini, Goldie, Yazgan, Jiang, Songhori, Wang, Lee, Johnson, Pathak, Nazi, et~al.]{mirhoseini2021graph}
Azalia Mirhoseini, Anna Goldie, Mustafa Yazgan, Joe~Wenjie Jiang, Ebrahim Songhori, Shen Wang, Young-Joon Lee, Eric Johnson, Omkar Pathak, Azade Nazi, et~al.
\newblock A graph placement methodology for fast chip design.
\newblock \emph{Nature}, 594\penalty0 (7862):\penalty0 207--212, 2021.

\bibitem[Schick et~al.(2023)Schick, Dwivedi-Yu, Dess{\`\i}, Raileanu, Lomeli, Hambro, Zettlemoyer, Cancedda, and Scialom]{schick2023toolformer}
Timo Schick, Jane Dwivedi-Yu, Roberto Dess{\`\i}, Roberta Raileanu, Maria Lomeli, Eric Hambro, Luke Zettlemoyer, Nicola Cancedda, and Thomas Scialom.
\newblock Toolformer: Language models can teach themselves to use tools.
\newblock \emph{Advances in Neural Information Processing Systems}, 36, 2023.

\bibitem[Sreedhar et~al.(2025)Sreedhar, Ogbonda, Yin, Shahidi, Nagaraj, Deng, Cohen, Kalker, Kumar, Yazdanbakhsh, and Subramanian]{sreedhar2025leveraging}
Kavya Sreedhar, Josh Ogbonda, Pengqi Yin, Narges Shahidi, Kanthi Nagaraj, Zhijie Deng, Rami Cohen, Ton Kalker, Sameer Kumar, Amir Yazdanbakhsh, and Suvinay Subramanian.
\newblock Leveraging {LLMs} to improve hardware-software co-design workflow productivity and accessibility.
\newblock In \emph{Workshop on Machine Learning for Computer Architecture and Systems (MLArchSys) at ISCA}, 2025.

\bibitem[Yao et~al.(2023)Yao, Zhao, Yu, Du, Shafran, Narasimhan, and Cao]{yao2023react}
Shunyu Yao, Jeffrey Zhao, Dian Yu, Nan Du, Izhak Shafran, Karthik Narasimhan, and Yuan Cao.
\newblock {ReAct}: Synergizing reasoning and acting in language models.
\newblock In \emph{International Conference on Learning Representations}, 2023.

\end{thebibliography}

\end{document}